\documentclass[11pt]{article}
\usepackage{eacl2017}
\usepackage{times}
\usepackage{url}
\usepackage{latexsym}
\usepackage{amsmath}
\usepackage{graphicx}
\usepackage{multirow}
\usepackage{url}
\usepackage{booktabs}
\usepackage{float}
\usepackage{styluscore}
\usepackage{stylusmath}
\usepackage{xcolor}
\usepackage{framed}
\usepackage{tablefootnote}
\usepackage{todonotes}
\usepackage{microtype}

\eaclfinalcopy

\begin{document}

  \title{Discriminative Information Retrieval for Knowledge Discovery}

   \author{Tongfei Chen \and Benjamin Van Durme \\
     Center of Language and Speech Processing \\
     Johns Hopkins University \\
     \Tt{\{tongfei,vandurme\}@cs.jhu.edu} \\
   }

  \maketitle

  \Abstract {
    We propose a framework for discriminative Information Retrieval (IR) atop linguistic features, trained to improve the recall of tasks such as answer candidate passage retrieval, the initial step in text-based Question Answering (QA).  We formalize this as an instance of linear feature-based IR \cite{Metzler:2007fy}, illustrating how a variety of knowledge discovery tasks are captured under this approach, leading to a 44\% improvement in recall for candidate triage for QA.}

\Section[Introduction] {

    Question Answering (QA) with textual corpora is typically modeled as first finding a candidate set of passages (sentences) that may contain an answer to a question, followed by an optional candidate reranking stage, and then finally an Information Extraction (IE) step to select the answer string.  QA systems normally employ an Information Retrieval (IR) system to produce the initial set of candidates, usually treated as a black box, bag-of-words process that selects candidate passages best overlapping with the content in the question. 

    \FigureStar[hbt] {
      \label{workflow}
      \Image[width=14cm]{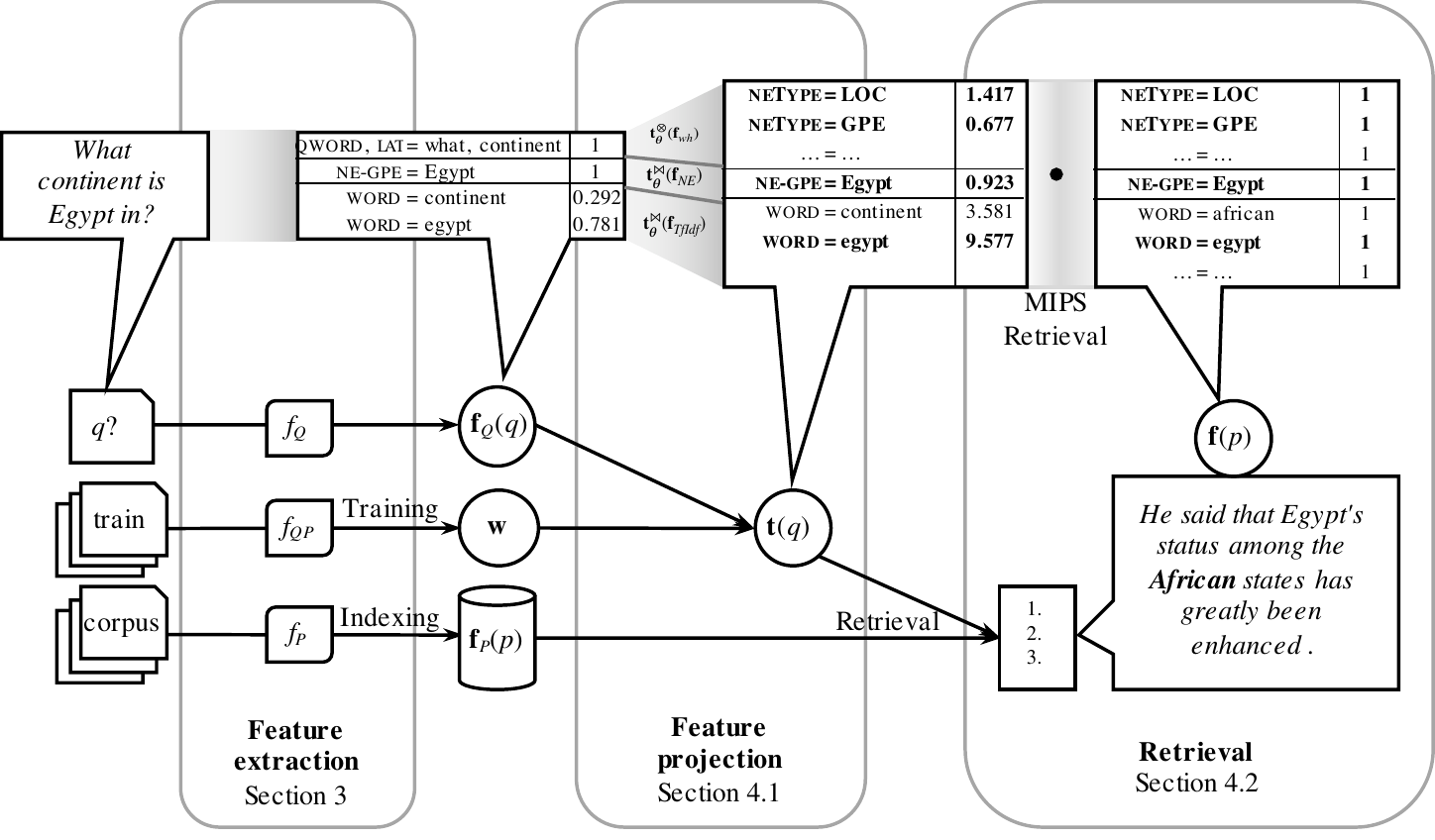}
      \centering
      \Caption{General workflow of DiscK (using question answering sentence retrieval as an example).}
    }

    Recent efforts in corpus-based QA have been focused heavily on reranking, or \emph{answer sentence selection}: filtering the candidate set as a supervised classification task to single out those that \emph{answer} the given question. Extensive research (see \S~Related Work) has explored employing syntactic/semantic features \cite{yih2013question,wang2010probabilistic,heilman2010tree,yao2013answer} and recently using neural networks \cite{severyn2015learning,wang2015long,yin2015abcnn}.  The shared aspect of all these approaches, no matter how state-of-the-art they may be, is that the quality of reranking a candidate set is upper bounded by the initial set of candidates.  Put another way: unless one plans on reranking the entire corpus (e.g., every sentence on the web) for each question as it arrives, one is still reliant on an initial IR stage in order to obtain  a computationally feasible QA system.

    We propose a framework called \It{DiscK} (Discriminative IR for Knowledge Discovery) for performing this triage step for QA sentence selection and other related tasks in \It{sublinear} time. 
    DiscK shows a log-linear model can be trained to optimize an objective function for downstream reranking, and the resulting trained weight can be reused to retrieve a candidate list.
     Our approach follows \newcite{yao2013automatic} who proposed the automatic coupling of answer sentence selection and information retrieval by augmenting a bag-of-words query with desired named entity (NE) types based on a given question. While Yao et al. showed improved performance in retrieval as compared with a bag-of-words baseline IR system, the model was proof-of-concept,  employing a simple linear interpolation between bag-of-words and NE features with a single scalar value tuned on a development set, kept static across all types of questions at test time.

     We improve upon Yao et al. in  two ways. First, we generalize their intuition by casting the problem  as an instance of classification-based retrieval \cite{robertson-spark-jones-76}, formalized as a discriminative (log-linear) retrieval model \cite{cooper-et-al-sigir-92,gey-sigir-94,nallapati-sigir-04} allowing for the use of rich NLP features.  Our framework can then be viewed as an instance of linear feature-based IR, following \newcite{Metzler:2007fy}.  Second, we illustrate how this approach equally applies across a variety of knowledge discovery tasks beyond QA, such as coreferential entity retrieval, and entity linking.

     Our experiments are based on a new, robust software package for structured feature extraction and retrieval,\footnote{Will be open-sourced released upon paper publication.} which shows state of the art performance in  QA sentence selection on the retrieval dataset of Lin and Katz (2006).



  }

  \Section[General Approach] {

    \Paragraph[Problem formulation] {


      Formally, given a candidate set $\calD = \{p_1, \cdots, p_N\}$, a query $q$ and a scoring function $F(q, p)$, an IR system retrieves the top-$k$ items under  the following objective:
\vspace{-2mm}
      \LabeledEq[DiscIRObjective] {
        \Argmax {p \in \calD} { F(q, p) }
      }

      Tackling Eq. (\ref{DiscIRObjective}) via straight-forward application of supervised classification (e.g., earlier mentioned, recent neural network based models) requires a traversal over all possible candidates, i.e. the corpus, which is computationally infeasible for any reasonable collection.
    }

    \Paragraph[Model overview] {

      Let $\bff_Q(q)$ refer to feature extraction on the query $q$, with corresponding candidate-side feature extraction $\bff_P(p)$ on the candidate, and finally $\bff_{QP}(q, p)$  extracts features from a (query, candidate) pair is defined in terms of $\bff_Q$ and $\bff_P$ via composition (defined later):
\vspace{-2mm}
      \LabeledEq[eqFeatureComposition] {
        \bff_{QP}(q, p) = C(\bff_Q(q), \bff_P(p))
      }

      From a set of query/candidate pairs we can train a model $M$ such that given the feature vector of a pair $(q, p)$, its returning value $M(\bff_{QP}(q, p))$ represents the predicted probability of whether the passage $p$ answers the question $q$. This model is chosen to be a log-linear model with the feature weight vector $\bftheta$, leading to the  optimization problem:
\vspace{-2mm}
      \begin{align}
      \label{opt}
        & \Argmax {p \in \calD} { M \Paren{ \bff_{QP}(q, p) } } \nonumber \\
        =& \Argmax{p \in \calD} {\bftheta \cdot \bff_{QP}(q, p) }
      \end{align}

      This is in accordance with the pointwise reranker approach, and is an instance of the linear feature-based model of \newcite{Metzler:2007fy}.

      Under specific compositional operations in $\bff_{QP}$, Eq. (\ref{opt}) can be transformed to:
\vspace{-2mm}
      \LabeledEq[TransformedOpt] {
        \Argmax{p \in \calD} {\bft_{\bftheta} (\bff_Q(q)) \cdot \bff_P(p) }
      }

      This is elaborated in \S~4. We project the original feature vector of the query $\bff_Q(q)$ to a transformed version $\bft_{\bftheta}(\bff_Q(q))$: this transformed vector is dependent on the model parameters $\bftheta$, where the association learned between the query and the candidate is incorporated into the transformed vector. This is a weighted, trainable generalization of \It{query expansion} in traditional IR systems.

       Under this transformation we observe that the joint feature function $\bff_{QP}(q, p)$ is decomposed into two parts with no interdependency -- the original problem in Eq. (\ref{opt}) is reduced to a standard \It{maximum inner product search} (MIPS) problem in Eq. (\ref{TransformedOpt}). Under sparse assumptions (where the query vector and the candidate feature vector are both sparse), this MIPS problem can be efficiently (sublinearly) solved using classical IR techniques (multiway merging of postings lists).


    }
    \Paragraph[Applications]{
 A variety of knowledge discovery tasks can be considered a ranking problem where the candidate set is large and where our general DiscK framework would be applicable, such as:

      \noindent\emph{Question answer sentence selection} {
         Given a natural language question $q$, and a set of candidate passages $\calD = \{p_1, \cdots, p_N\}$ (all the sentences in the given corpus), the model ranks $p_i$, i.e. retrieves the top-$k$ passages wrt how well they provide an answer to the question $q$.
      }

      \noindent\emph{Dynamic cross-document coreference resolution} {
        Given an entity mention $m$ in a sentence/document, and a set of candidate mentions $\calD = \{m_1, \cdots, m_N\}$, the model ranks $m_i$, i.e. retrieves the top-$k$ mentions wrt how possible they are coreferential to $m$.
      }

     \noindent\emph{Slot filling} {
       Given an entity mention $m$ and a relation $R$, and a set of candidate mentions $\calD = \{m_1, \cdots, m_N\}$ (all entities discovered by an in-document coreference system in the given corpus), the model ranks $m_i$, i.e. retrieves the top-$k$ entities wrt how probable the expression $R(m, m_i)$ holds.
     }

     \noindent\emph{Entity linking} {
       Given an entity mention $m$, and a set of candidate entities $\calD = \{e_1, \cdots, e_N\}$ (e.g. all articles in Wikipedia or all entities in Freebase), the model ranks $e_i$, i.e. retrieves the top-$k$ possible entities with respect to how probable the mention $m$ is linked to entity $e_i$.
     }
   }
  }
  \Section[Features] {

    This section describes the feature engineering framework used in DiscK, using question answer sentence selection as the motivating task.  A feature vector can be seen as an associative array that maps  features in the form ``\Sc{key}=\It{value}'' to real-valued weights. One item in a feature vector $\bff$ is denoted as ``$ (\Sc{key}=\It{value}, \It{weight}) $'', and a feature vector can be seen as a set of such tuples. We write $\bff_{(\Sc{key}=\It{value})} = \It{weight}$ to indicate that the features serve as keys to the associative array, and $\theta_{X}$ is the weight of the feature $X$ in the trained model $\bftheta$.  We set the constraint that all candidate features are binary (their weights are exclusively 0 or 1).

  \Subsection[Features for questions and passages] {

    Features used for a question are listed as follows.

             \noindent \It{\Bf{Question word}} ($\bff_{\It{wh}}$):
              The type of the question, typically the \It{wh}-word of a sentence. If it is a question like ``\It{How many}'', the word after the question word is also included in the feature, i.e., feature ``(\Sc{qword}=\It{how many}, 1)'' will be added to the feature vector of the question.

            \noindent \It{\Bf{Lexical answer type}} ($\bff_{\It{lat}}$):
             If the query is a question where the question word is ``what'' or ``which'', we identify the lexical answer type (LAT) of this question \cite{ferrucci2010building}, which is defined as the head word of the first NP after the question word. For example, the LAT feature from the question ``\It{What is the city of brotherly love?}'' would be ``(\Sc{lat}=\It{city}, 1)''. If the question word is not ``what'' or ``which'', generate an empty feature (\Sc{lat}=$\emptyset$, 1).

             \noindent \It{\Bf{Named entities}} ($\bff_{\It{NE}}$):
       All the named entities discovered in this question. For example, features like ``(\Sc{ne-person}=\It{Margaret Thatcher}, 1)'' would be added to the feature vector of the question if Thatcher is mentioned in the sentence. 

             \noindent \It{\Bf{Normalized \It{tf-idf} weighted bag-of-words features}} ($\bff_{\It{TfIdf}}$):
       The $L_2$-normalized \It{tf-idf} weighted bag-of-words feature of this question. An example feature would be ``(\Sc{word} = \It{author}, 0.454)''.

      The features used for candidate passages are listed as follows, constrained to be binary.

           \noindent \It{\Bf{Bag of words}} ($\bff_{\It{BoW}}$): Any distinct word \It{x} in the passage will generate a feature ``(\Sc{word}=\It{x}, 1)''.

           \noindent \It{\Bf{Named entity types}} ($\bff_\It{NEType}$): Types of named entities. For example, If the passage contains a name of a person, a feature ``(\Sc{ne-type}=PERSON, 1)'' will be generated.

           \noindent \It{\Bf{Named entities}} ($\bff_\It{NE}$): All the named entities discovered in this passage. This is the same feature as the named entity feature for questions.
    }

    \Subsection[Feature vector composition] {
    \label{secComposition}

    This section elaborates the composition $C$ of the question feature vector and passage feature vector.

    We define two operators on feature vectors: Cartesian product ($\otimes$) and join ($\bowtie$).

    For any feature vector of a question $\bff_Q(q) = \{ (k_i = v_i, w_i) \}, (w_i \le 1)$\footnote{If $w_i > 1$, the vector can always be normalized so that the weight of every feature is less than 1.} and any feature vector of a passage $\bff_P(p) = \{ (k_j = v_j, 1) \}$, the Cartesian product of them is defined as
    \Eq{
    \bff_Q(q) \otimes \bff_P(p) = \{ ( (k_i, k_j) = (v_i, v_j), w_i) \};
    }
    whereas the join of them is defined as
    \Eq{
    \bff_Q(q) \bowtie \bff_P(p) = \{ ((k_i=k_j) = 1, w_i) \}.
    }

    Notation $(k_i=k_j)=1$ denotes that the value for feature $k_i$ on the question side is the same as the feature $k_j$ on the passage side.
    
    The composition that generates the feature vector for the question/passage pair in Eq. (\ref{eqFeatureComposition}) is therefore defined as
    \begin{align}
      &\begin{matrix*}[l]
      C(& \bff_Q(q) &,& \bff_P(p) &)  \\
      = &({{{\bff}_{\It{wh}}}(q)}\otimes\bff_{\It{lat}}(q))    &\otimes &{{{\bff}_{\It{NEType}}}(p)}& \\
      + & (\bff_{\It{wh}}(q)\otimes\bff_{\It{lat}}(q))         &\otimes & \bff_{\It{BoW}}(p)& \\
      + &{{{\bff}_{\It{NE}}}(q)}     &\bowtie &{{{\bff}_{\It{NE}}}(p)}& \\
      + &{{{\bff}_{\It{TfIdf}}}(q)}  &\bowtie &{{{\bff}_{\It{BoW}}}(p)}&.
      \end{matrix*}
    \end{align}

    $(\bff_{\It{wh}}(q) \otimes \bff_{\It{wh}}(q)) \otimes \bff_{\It{NEType}}(p)$ captures the association of question words and lexical answer types with the expected type of named entities. During training, we discovered features like (\Sc{qword,ne-type})=(who, PERSON), (\Sc{qword,ne-type})=(when, DATE), (\Sc{lat,ne-type})=(city, GPE\footnote{GPE: Geo-political entities.}) will be assigned high weights.

     $(\bff_{\It{wh}}(q) \otimes \bff_{\It{wh}}(q)) \otimes \bff_{\It{BoW}}(p)$ captures the relation between some question types with certain words in the answer. For example, we observed feature ``$\rm (\Sc{(qword,lat),word)}=(what,capacity),gallon)$'' to have a relative high weight, because the word ``gallon'' can be expected from a question asking about capacity.

    ${{{\bff}_{\It{NE}}}(q)}\bowtie {{{\bff}_{\It{NE}}}(p)}$ captures named entity overlap. Features like $(\Sc{ne-person=ne-person})=1$ will be assigned high weights because sentences talking about the same person will have high question-answer association. Interestingly, we observed feature $(\Sc{ne-norp}\footnote{NORP: Nationality.}=\Sc{ne-language})=1$ is of high weight, because words like ``French'' can refer to either a language or an adjective meaning ``pertaining to France''. This kind of feature helps mitigates the error of named entity annotations.\footnote{As observed by Yao et al.}

    ${{{\bff}_{\It{TfIdf}}}(q)}\bowtie {{{\bff}_{\It{BoW}}}(p)}$ measures general \It{tf-idf}-weighted context word overlap. Using only this feature without the others effectively reduces the system to a traditional \It{tf-idf}-based retrieval system.

    }

  }

  \Section[Feature Projection and Retrieval] {

    \Subsection[Feature Projection] {
    
      Given a question, it is desired to know what kind of features that its potential answer might have. Once this is known, an index searcher will do the work to retrieve the desired passage.

      According to the feature composition method in Section \ref{secComposition}, there are two ways of composition: Cartesian product and join.

      For Cartesian product, we define

      \Eq{
        \bft_{\bftheta}^{\otimes}(\bff) = \{(k^\prime = v^\prime, w \theta_{(k, k^\prime) = (v, v^\prime)}) | (k=v, w) \in \bff \},
      }
      for all $k^\prime, v^\prime$ such that $\theta_{(k, k^\prime) = (v, v^\prime)} \ne 0$, i.e. feature $(k, k^\prime) = (v, v^\prime)$ appears in the trained model.

      Take the example in Fig. \ref{workflow} as an example. In the feature vector of the query a feature $(\Sc{qword,lat})=(\Rm{what,continent})$ is present. Because of the presence of the feature $(\bff_{wh} \otimes \bff_{lat}) \otimes \bff_{\It{NEType}}$, this feature vector will be projected to a feature vector of named entity types.

      For join, we have

      \Eq{
        \bft_{\bftheta}^{\bowtie}(\bff) = \{(k^\prime = v, w \theta_{(k= k^\prime) = 1}) | (k=v, w) \in \bff \},
      }
      for all $k^\prime$ such that $\theta_{(k = k^\prime) = 1} \ne 0$, i.e. feature $(k = k^\prime) = 1$ appears in the trained model.

      Consider the example in Fig. \ref{workflow}: the query feature vector contains $(\Sc{ne-gpe}=\Rm{Egypt})$. Because of the presence of the feature vector $\bff_{NE} \bowtie \bff_{NE}$, this feature vector will be also projected to a feature vector of named entities. The projected feature vector is shown in Fig. \ref{workflow}.

      It can be shown from the definitions above that:
      \begin{align}
        \bft_{\bftheta}^{\otimes}(\bff) \cdot \bfg &=  \bftheta \cdot (\bff \otimes \bfg) ; \\
        \bft_{\bftheta}^{\bowtie}(\bff) \cdot \bfg &=  \bftheta \cdot (\bff \bowtie \bfg) .
        \end{align}

      The transformed feature vector $\bft(q)$ of an expected answer passage given a feature vector of a question $\bff_Q(q)$ is:
      \begin{align}\label{eqPrediction}
        \bft(q) =
        \bft_{\bftheta}^\otimes(\bff_\textit{wh}(q) \otimes \bff_\It{lat}(q)) +
        \bft_{\bftheta}^{\bowtie} (\bff_\textit{NE}(q) + \bff_{\textit{TfIdf}}(q)).
      \end{align}

      Calculating the vector $\bft(q)$ is computationally efficient because it only involves sparse vectors.

      Finally,  our initial optimization problem stated in Eq. (\ref{opt}) is then  equivalent to:
      \begin{align}\label{eqDotProductObjective}
        & \Argmax{p \in \mathcal{D}} {\bftheta \cdot C(\bff_Q(q), \bff_P(p))} \nonumber \\
        = & \Argmax{p \in \mathcal{D}} {\bft_{\bftheta}(q) \cdot \bff_P(p)}.
      \end{align}

      Succinctly, we have just formally shown that given a question, we can reverse-engineer the features we expect to be present in a candidate using the transformation function $t$, which we will then use a query vector for retrieval.

    }

  \Subsection[Retrieval] {

    We use Apache \Sc{Lucene}\footnote{\texttt{http://lucene.apache.org}.} to build the index of the corpus, which, in the scenario of DiscK, is the feature vectors of all candidates $\bff_P(p), p \in \calD$. This is an instance of weighted bag-of-features instead of common bag-of-words.

    For a given question $q$, we first compute its feature vector $\bff(q)$ and then compute its transformed feature vector $\bft_{\bftheta}(q)$ given model parameters $\bftheta$. Together this forms a weighted query to the \Sc{Lucene} index. We modified the similarity function of \Sc{Lucene} when executing multiway postings list merging so that fast efficient maximum inner product search can be achieved. This classical IR technique ensures sublinear performance because only vectors with at one overlapping feature, instead of the whole corpus, is traversed.
    \footnote{The closest work of indexing we are aware of is Bilotti et al. \cite{bilotti2007structured}, which transformed linguistic structures to structured constraints, which is different from DiscK's approach of directly indexing linguistic features.}
    }
  }

  \Section[Experiments] {

    \Subsection[Question Answering Sentence Retrieval] {

      \Paragraph[Data and setup] {

        We use the training and test data from Yao et al. \shortcite{yao2013automatic}. The corpus from which the passages are retrieved is the AQUAINT Corpus of English News Text \cite{aquaint}. All sentences with more than 100 tokens are discarded. In this dataset each question is paired with 10 answer candidates, which are retrieved from the whole AQUAINT corpus by using a vanilla maximum \It{tf-idf} cosine similarity search. For the 10 answer candidates for each question, whether it answers the given question is human annotated, thus serving as the gold truth for the training set.

        The test data is a set of questions from Lin and Katz \shortcite{lin2006building} with those that do not have an answer by matching the TREC answer patterns removed. The resulting data set consists of 99 questions.

        All corpora are NER-tagged by the Illinois Named Entity Tagger \cite{ratinov2009design} with an 18-label entity type set. Questions are parsed via the Stanford Parser \cite{klein2003accurate}.

        We divided the training data into two parts: 53 questions whose answer sentences can be found in the AQUAINT corpus are used as the development set and the rest is used for training the log-linear model. Very few sentences in the corpus provide an answer to a given query, i.e., most sentences in the corpus are negative examples.  We follow \newcite{nallapati-sigir-04} and undersample the majority (negative) class, taking 50 sentences uniformly at random from the AQUAINT corpus, per query, filtered to ensure no such sentence matches a query's answer pattern as additional negative samples to the training set. The summary of the datasets are shown in Table (\ref{tab:data-summary}) \footnote{The number of negative samples of the training set does not include the randomly sampled 50 negative samples for each training question.}.

        \begin{table}[H]
          \small
          \centering
          \caption{Summary of the datasets.}
          \label{tab:data-summary}
          \begin{tabular}{llll}
            \toprule
                                   & train & dev & test \\
            \midrule
            \# of questions        & 2150  & 53  & 99   \\
            \# of positive samples & 7421  & 216 & 368  \\
            \# of negative samples & 14072 & \multicolumn{2}{c}{23 million +}  \tablefootnote{This is the total number of sentences in the corpus from which answers are to be retrieved.}  \\
            \bottomrule
          \end{tabular}
        \end{table}

      }

    }

The model is trained using \textsc{Liblinear} \cite{fan2008liblinear}, with heavy $L_1$-regularization to the maximum likelihood objective. The regularization coefficient is tuned on a dev set, with the objective of maximizing mean average precision (MAP).

    \Paragraph[Baseline systems] {

      We include the following IR systems as baselines for comparison. We stress that recent prior work in  neural network based \emph{reranking} is not directly applicable here as those runtimes are \It{linear} with respect to the number of candidate sentences, which is computationally infeasible given a large corpus.

      \noindent\textit{Off-the-shelf \textsc{Lucene}}: Directly indexing the sentences in \textsc{Lucene} and do sentence retrieval. This is equivalent to maximum \textit{tf-idf} retrieval.

      \noindent\textit{Yao et al. (2013b)}: A  retrieval system which augments the bag-of-words query with desired named entity types based on a given question.
    }

    \Paragraph[Evaluation metrics] {
    (1) \Bf{R@1k}: The recall in top-1000 retrieved list. Contrary to normal IR systems which optimize precision (as seen in metrics such as P@5 and P@10), our system is a triaging system whose goal is to retrieve good candidates for downstream reranking. So we chose the R@1k as a metric to measure the quality of the triaging. (2) \Bf{b-pref}: The b-pref measure \cite{buckley2004retrieval} is designed for situations where relevance judgments are known to be far from complete. It computes a preference relation of whether judged relevant documents are retrieved ahead of judged irrelevant documents. (3) \Bf{MAP}: mean average precision and (4) \Bf{MRR}: mean reciprocal rank.
    }

    \Paragraph[Results] {

\begin{table}[H]
\small
\centering
\caption{Performance of the QA retrieval systems.}
\label{tab:performance}
\begin{tabular}{lrrrr}
\toprule
                      & R@1k  & b-pref  & MAP     & MRR \\
\midrule
\color{gray}{Lucene (dev)}  & \color{gray}{28.65\%}  & \color{gray}{40.46\%} & \color{gray}{9.63\%}  & \color{gray}{13.94\%} \\
Lucene (test) & 33.76\%  & 31.89\% & 9.78\%  & 15.06\% \\
Yao+ (test)\tablefootnote{Results on dev data is not reported.} & 25.88\% & 45.41\% & 13.75\% & \Bf{29.87\%} \\
\color{gray}{DiscK (dev)}         & \color{gray}{71.64\%} & \color{gray}{70.65\%} & \color{gray}{20.41\%}  & \color{gray}{30.09\%} \\
DiscK (test)        & \bf 77.99\% & \bf 75.24\% & \bf 16.68\%  & 22.21\% \\
\bottomrule
\end{tabular}
\end{table}

      Our method performs significantly better than Yao et al., demonstrating the effectiveness of trained weighted queries compared to binary augmented features.  The performance gain with respect to off-the-shelf \textsc{Lucene} with reranking shows that the our weighted augmented queries by decomposition is superior to vanilla \textit{tf-idf} retrieval. The significantly higher recall and b-pref score of discriminative IR shows that our proposed method results in much better top-$k$ triage. Downstream rerankers (e.g. neural networks) should benefit in the future from the improved triaging provided by DiscK.
    }

    \Paragraph[Error analysis] {

      We list some typical negative samples of the DiscK system running over the question answering sentence retrieval tasks. Most of these errors arise from the lack of expressiveness of the feature functions, as can be shown below:

      \begin{framed}\small
      Q: \It{What is the abbreviation of London Stock Exchange?}

      A: \It{The London Stock Exchange (\Bf{LSE}) board Thursday agreed to introduce its controversial computerized order-driven share dealing system, but not for at least another year and initially only for FT-SE 100 stocks.}
      \end{framed}
      Failed because the current feature set is unable to capture how to answer an ``abbreviation'' question. If we add an additional lightweight feature to detect all-caps words in the feature side ($\bff_{\It{allCaps}}$ will fire if there exists an all-caps word in the candidate sentence), and add a feature $(\bff_{wh}\otimes\bff_{lat})\otimes\bff_{\It{allCaps}}$, given enough training data, a feature $(\Sc{((qword,lat),allCaps)}=(\Rm{(what,abbreviation), TRUE})$ will probably gain a high weight in the training process, hence enabling the system to correctly answer this instance of question.

%

      \begin{framed}\small
        Q: \It{What is the fastest car in the world?}

        A: \It{The \Bf{Thrust SuperSonic Car} set a world land speed record of 1,142 kph on September 25.}
      \end{framed}
      DiscK failed to retrieve this sentence because of the inability of the feature set to learn the association between ``fastest'' and ``speed record''. This is best solved by distributed semantic representation of words and sentences, which are inherently dense, real-valued vectors. Because DiscK relies on sparsity to achieve its sublinear retrieving performance, these dense features are out of scope for DiscK.

    }

    \Subsection[Cross-document Coreferent Mention Retrieval] {
      To illustrate the applicability of DiscK to other corpus knowledge discovery tasks, we provide a proof-of-concept formulation for searching for mentions that are coreferent to a query mention in some given document.
      \Paragraph[Data and setup] {
        We use the TAC KBP 2014 English Entity Discovery and Linking (EDL) data \footnote{LDC2014E54 for training and development and LDC2014E81 as the test data. The base corpus for all the text is LDC2014E13. These datasets can be found at \Tt{http://www.nist.gov/tac/2014/KBP/data.html}.} as our corpus. The corpus comprises of text from English Gigaword, discussion forums and webposts. This is an entity-linked corpus in which some of the mentions in the text is labeled with a grounded entity ID. If two mentions point to the same entity, then they are coreferent mentions.  For each unique entity, which has a set of mentions $\{m_1, \cdots, m_{n}\}$ linked to it, we select one mention $m_1$ as the query, and the rest are relevant candidates. Our cross-doc coreference resolution system is supposed to find these coreferent mentions $\{m_2, \cdots, m_n\}$ when given the query mention $m_1$. Entities that with only one linking mention are discarded.

        All corpora are NER-tagged using Stanford CoreNLP \cite{manning-EtAl:2014:P14-5}. All named entities discovered by the Stanford NER will be the index that our system is to retrieve from.  We divide the training corpus (LDC2014E54) into two parts with no overlapping documents: one for training and one for development. In the training set, all coreferent mention pairs that are not in the same document are extracted as positive samples, and additional 50 negative samples are uniformly randomly sampled for each mention. The testing corpus is LDC2014E81. The summary of the datasets is shown in Table (\ref{tab:xdoc-data-summary}).

        \begin{table}[H]
          \small
          \centering
          \caption{Summary of Dynamic Cross-Doc. Coref. datasets.}
          \label{tab:xdoc-data-summary}
          \begin{tabular}{llll}
            \toprule
                                    & train & dev & test \\
            \midrule
            \# of distinct entities & 802   & 192  & 721  \\
            \# of labeled mentions  & 4059  & 704  & 3379 \\
            \# of all mentions to be retrieved & \multicolumn{3}{c}{100 million +} \\
            \bottomrule
          \end{tabular}
        \end{table}

        The training process of the model follows the aforementioned process in the QA task.
      }

      \Paragraph[Features] {
        We designed the following feature composition function for mention-entity pairs.

        \noindent\Bf{\It{Mention text}} ($\bff_{text}$): The text of an entity mention itself.

        \noindent\Bf{\It{Type}} ($\bff_{type}$): The type of an entity mention as determined by the NER system. It could be PERSON, LOC, GPE, ORG or others.

        \noindent\Bf{\It{Acronym}} ($\bff_{acro}$): The acronym of the text of an entity mention. For example, the entity mention ``United States of America'' yields a feature (\Sc{acro}=USA, 1).

        \noindent\Bf{\It{Letter trigrams}} ($\bff_{l3g}$): The letter trigrams of each word in the entity mention. For instance, the mention ``Tehran'' yields 4 features: (\Sc{l3g}=teh, 1), (\Sc{l3g}=ehr, 1), (\Sc{l3g}=hra, 1) and (\Sc{l3g}=ran, 1). This feature function is designed to mitigate the problem of different Romanizations of names originally written in non-English scripts.

        \noindent\Bf{\It{Document context}} ($\bff_{dc}$): Extracts the top-10 hightest \It{tf-idf} words in the surrounding document of the query being queried. This helps the system identify the topic of context of the query.

        For each entity mention $m$, we have
        \LabeledEq[xdcFeatures] {
          \bff_{overall} = \bff_{type} \otimes (\bff_{text} + \bff_{acro} + \bff_{l3g} + \bff_{dc}).
        }

        The features like $\bff_{acro}$ or $\bff_{l3g}$ are sometimes related to the mention types. For example, the acronym of organizations (e.g. United States of America, USA), and the letter trigrams of names of people are useful but the acronym of locations are probably not so useful. To take this type-dependent information into account, the type of an entity mention is paired with every other feature, as shown in the Cartesian product operation in Eq. (\ref{xdcFeatures}). Additionally, this can help us eliminate most candidates whose types are different from our query.

         In the case of cross-document coreference, the query and the candidate are symmetric: the features are both sides are the same. Henceforth the pairwise feature function is

        \begin{align}
          \begin{matrix*}[l]
           C(& \bff_Q(q) &,& \bff_P(p) &)  \\
           = &{{{\bff}_{\It{overall}}}(q)}\    &\bowtie &{{{\bff}_{\It{overall}}}(p)}&.
          \end{matrix*}
        \end{align}

      }

      \Paragraph[Evaluation metrics] {
        We reuse those metrics as in the previous task. Because of the potential large number of coreferent mentions of specific mentions in the set, the recall metric R@1000 is changed to R@10000. Additionally, because the annotations of the coreferent mentions are far from complete (e.g. only a very small portion of coreferent mentions of ``Britain'' are annotated in the corpus), b-pref, designed for situations where only partial relevance judgements exist, is especially suitable here.
      }

      \Paragraph[Results] {

      \begin{table}[H]
          \small
          \centering
          \caption{Performance of entity mention search.}
          \label{tab:xdoc-performance}
          \begin{tabular}{lrrrr}
          \toprule
                      & R@10k  & b-pref  & MAP     & MRR \\
          \midrule
          String match & 10.53\%  & 11.44\% & 2.16\%  & 0.96\% \\
          DiscK & \bf 27.13\% & \bf 33.27\% & \bf 9.02\%  & \bf 5.53\% \\
          \bottomrule
          \end{tabular}
      \end{table}

      }

     Direct string matching is used for comparison: if two mentions have the same text, this system judges them as coreferent, otherwise not.

      The results are shown in Table \ref{tab:xdoc-performance}, with DiscK clearly outperforming basic string match by a large margin in all evaluation metrics.  We do not here seek to establish  competitiveness against prior related work in cross-document coreference,\footnote{E.g., the work of \newcite{bagga1998entity}, \newcite{gooi2004cross}, \newcite{mayfield2009cross}, \newcite{culotta2006first}, \newcite{poon2008general}, \newcite{wick2009entity}, or \newcite{singh2011large}.} only to illustrate that this task could be viewed as an application the DiscK framework.  Future work may consider further experimentation, especially when paired with a high-performance discriminative reranking model.

    }

  }

  \Section[Related Work] {

    \Paragraph[Discriminative Information Retrieval] {

      Traditional information retrieval (IR) models viewed the retrieval problem as measuring the similarity, often the cosine similarity between two bag-of-words vectors, between the query and the candidates. One shortcoming of this vector-space model (VSM) is that it did not provide a theoretical basis for computing the optimum weights. The binary independence retrieval (BIR) \cite{robertson-spark-jones-76} viewed IR as a classification problem that classifies the entire collection of candidates into two classes: relevant and irrelevant. In the framework of BIR, a probability of $P(p, q)$ is computed and ranked to generate the retrieved list. In this view of casting IR as a discriminative model, sophisticated machine learning techniques can be leveraged.

      Cooper et al. \shortcite{cooper-et-al-sigir-92}, Gey \shortcite{gey-sigir-94}, and Nallapati \shortcite{nallapati-sigir-04} further formalized this framework into a logistic regression (log-linear) retrieval model. Another prominent example of employing discriminative models in IR is by language modeling \cite{ponte1998language}.

    }

    \Paragraph[Question Answering Sentence Selection] {
      There exists substantial previous work on question answering sentence selection, or more generally, sentence pair modeling.

      \noindent\emph{Syntactic and Semantic Analysis} {
      Bag of words representation with simple surface form matching often results in poor predictive power, leading to prior work exploring syntactic and semantic structures of the text. Bilotti et al. \shortcite{bilotti2007structured} preprocessed the corpus with a semantic parser and an NER system. These semantic analyses are expressed as structural constraints on semantic annotations and keywords, and are translated directly into structured queries. Moldovan et al. \shortcite{moldovan2007cogex} transformed questions to logic representations based on their syntactic, semantic and contextual information, utilizing a logic prover to perform QA. Punyakanok et al. \shortcite{punyakanok2004mapping}, Heilman and Smith \shortcite{heilman2010tree} and Yao et al. \shortcite{yao2013answer} used tree edit distance, and Wang et al. \shortcite{wang2007jeopardy} employed quasi-synchronous grammars to match the dependency parse trees of the question and the answer sentence.
      }

      \noindent\emph{Lexical semantic features} {
      Instead of utilizing higher-level abstractions such as syntactic and semantic analysis, another thread of previous work focussed on shallow lexical semantic features. Yih et al. \shortcite{yih2013question} performed semantic matching based on a latent word-alignment structure arising from WordNet. Lai and Hockenmaier \shortcite{lai2014illinois} utilized word relations such as words being synonyms, antonyms, hypernyms and hyponyms to perform a more fine-grained semantic overlap between sentences.
      }

      \noindent\emph{Neural methods} {
        Yu et al. \shortcite{yu2014deep} and Severyn and Moschitti \shortcite{severyn2015learning} proposed the use of  convolutional neural networks (CNNs) to model question and answer pairs, followed by Yang et al. \shortcite{yang2015wikiqa} with a related model and the introduction of the WikiQA dataset.\footnote{WikiQA is akin to the pre-existing dataset of Yao et al., but unlike Yao et al., WikiQA foregoes the connection to years of prior work in TREC-based QA evaluations: we therefore do not consider it here.} Tan et al. \shortcite{tan2015lstm} and Wang and Nyberg \shortcite{wang2015long} made use  of bidirectional LSTM networks to model question answer pairs. To better capture the interdependency between the question answer sentence pairs, Yin et al. \shortcite{yin2015abcnn} proposed a generic attention-based CNN to model the sentence pairs for question answering, paraphrase identification and textual entailment. Amiri et al. \shortcite{amiri2016learning} presented a pairwise context-sensitive autoencoder to computing text pair similarity, and achieved state-of-the-art performance on answer reranking.  All of these efforts were aimed at candidate set re-ranking, once an initial retrieval step had been performed.  Huang et al. \shortcite{Huang:2013fz} proposed to embed questions and the sentences of a provided corpus together into a shared vector space, followed by an ``argmax'' operation at query time to seek the sentence maximizing cosine similarity: they give no details on what is by default a linear operation in the size of the corpus, which is impractical for large collections as compared to our sub-linear retrieval approach.
      }
    }






  }

  \Section[Conclusion and Future Work] {

\newcite{yao2013automatic} proposed to couple information retrieval
with features from downstream question answer sentence selection.  We
generalized this intuition by recognizing it as an instance of
discriminative retrieval, and proposed a new framework, DiscK, for generating
weighted, feature-rich queries based on a given query (may be a natural language question or a mention as we discussed in this paper). This approach allows for the straightforward use of a
downstream model in the candidate selection process, and leads to a
significant gain in recall, b-pref and MAP in the triaging step compared to prior work, hence providing better candidates for downstream reranking models, which could be coupled to this approach in future work.

Our framework is general and should apply to a variety of other
structurally related tasks.  We release a software library that
implements our described feature abstraction and query generation:
future work might extend this for other information extraction (IE) tasks such as entity linking
(retrieving candidate entities from a large knowledge base given a query mention) and slot filling (retrieving candidate mentions from a large text corpus given a query mention and a relation).

}


  \bibliography{eacl2017}
  \bibliographystyle{eacl2017}

\end{document}